\documentstyle[12pt]{article}

\def\doit#1#2{\ifcase#1\or#2\fi}

\skewchar\fivmi='177 \skewchar\sixmi='177 \skewchar\sevmi='177
\skewchar\egtmi='177 \skewchar\ninmi='177 \skewchar\tenmi='177
\skewchar\elvmi='177 \skewchar\twlmi='177 \skewchar\frtnmi='177
\skewchar\svtnmi='177 \skewchar\twtymi='177
\def\@magscale#1{ scaled \magstep #1}

\def\framingfonts#1{
\doit{#1}{\font\twfvmi  = ammi10   \@magscale5 
\skewchar\twfvmi='177 \skewchar\fivsy='60 \skewchar\sixsy='60
\skewchar\sevsy='60 \skewchar\egtsy='60 \skewchar\ninsy='60
\skewchar\tensy='60 \skewchar\elvsy='60 \skewchar\twlsy='60
\skewchar\frtnsy='60 \skewchar\svtnsy='60 \skewchar\twtysy='60
\font\twfvsy  = amsy10   \@magscale5 
\skewchar\twfvsy='60
\font\go=font018			
\font\sc=font005			
\def\Go#1{{\hbox{\go #1}}}	
\def\Sc#1{{\hbox{\sc #1}}}	
\def\Sf#1{{\hbox{\sf #1}}}	
\font\oo=circlew10	      
\font\ooo=circle10			
\font\ro=manfnt				
\def\kcl{{\hbox{\ro 6}}}		
\def\kcr{{\hbox{\ro 7}}}		
\def\ktl{{\hbox{\ro \char'134}}}	
\def\ktr{{\hbox{\ro \char'135}}}	
\def\kbl{{\hbox{\ro \char'136}}}	
\def\kbr{{\hbox{\ro \char'137}}}	
}}

\catcode`@=11
\def\un#1{\relax\ifmmode\@@underline#1\else
	$\@@underline{\hbox{#1}}$\relax\fi}
\catcode`@=12

\let\du=\d			

\def\a{\alpha} \def\b{\beta}  \def\d{\delta}
\def\e{\epsilon}  \def\g{\gamma}
   
\def\l{\lambda} \def\m{\mu} \def\n{\nu} 
  \def\r{\rho} \def\s{\sigma}
   
  \def\G{\Gamma} 
\def\L{\Lambda}

\def\bo{{\raise.15ex\hbox{\large$\Box$}}}		
\def\pa{\partial}					
\def\pr{\prod}						
\def\TH{{\raise.2ex\hbox{$\displaystyle \bigodot$}\mskip-4.7mu \llap H \;}}
\def\face{{\raise.2ex\hbox{$\displaystyle \bigodot$}\mskip-2.2mu \llap {$\ddot
	\smile$}}}					

\def\sp#1{{}^{#1}}				
\def\Tilde#1{{\widetilde{#1}}\hskip 0.03in}			
\def\Hat#1{\widehat{#1}}			
\def\Bar#1{\overline{#1}}			
\def\leftrightarrowfill{$\mathsurround=0pt \mathord\leftarrow \mkern-6mu
	\cleaders\hbox{$\mkern-2mu \mathord- \mkern-2mu$}\hfill
	\mkern-6mu \mathord\rightarrow$}
\def\dvec#1{\vbox{\ialign{##\crcr
	\leftrightarrowfill\crcr\noalign{\kern-1pt\nointerlineskip}
	$\hfil\displaystyle{#1}\hfil$\crcr}}}		

\def\frac#1#2{{\textstyle{#1\over\vphantom2\smash{\raise.20ex
	\hbox{$\scriptstyle{#2}$}}}}}			
\def\sfrac#1#2{{\vphantom1\smash{\lower.5ex\hbox{\small$#1$}}\over
	\vphantom1\smash{\raise.4ex\hbox{\small$#2$}}}}	
\def\bfrac#1#2{{\vphantom1\smash{\lower.5ex\hbox{$#1$}}\over
	\vphantom1\smash{\raise.3ex\hbox{$#2$}}}}	
\def\afrac#1#2{{\vphantom1\smash{\lower.5ex\hbox{$#1$}}\over#2}}    

\newskip\humongous \humongous=0pt plus 1000pt minus 1000pt
\def\caja{\mathsurround=0pt}
\def\eqalign#1{\,\vcenter{\openup2\jot \caja
	\ialign{\strut \hfil$\displaystyle{##}$&$
	\displaystyle{{}##}$\hfil\crcr#1\crcr}}\,}
\newif\ifdtup
\def\panorama{\global\dtuptrue \openup2\jot \caja
	\everycr{\noalign{\ifdtup \global\dtupfalse
	\vskip-\lineskiplimit \vskip\normallineskiplimit
	\else \penalty\interdisplaylinepenalty \fi}}}
\def\li#1{\panorama \tabskip=\humongous				
	\halign to\displaywidth{\hfil$\displaystyle{##}$
	\tabskip=0pt&$\displaystyle{{}##}$\hfil
	\tabskip=\humongous&\llap{$##$}\tabskip=0pt
	\crcr#1\crcr}}

\doit0{
\def\ref#1{$\sp{#1)}$}
}

\parskip=\medskipamount			
\thicklines			    

\def\border{						
	\setlength{\unitlength}{1mm}
	\newcount\xco
	\newcount\yco
	\xco=-24
	\yco=12
	\begin{picture}(140,0)
	\put(\xco,\yco){$\ktl$}
	\advance\yco by-1
	{\loop
	\put(\xco,\yco){$\kcl$}
	\advance\yco by-2
	\ifnum\yco>-240
	\repeat
	\put(\xco,\yco){$\kbl$}}
	\xco=158
	\yco=12
	\put(\xco,\yco){$\ktr$}
	\advance\yco by-1
	{\loop
	\put(\xco,\yco){$\kcr$}
	\advance\yco by-2
	\ifnum\yco>-240
	\repeat
	\put(\xco,\yco){$\kbr$}}
        \put(-20,11){\tiny University of Maryland Elementary Particle
Physics University of Maryland Elementary Particle Physics University of
Maryland Elementary Particle Physics}
	\put(-20,-241.5){\tiny University of Maryland Elementary
Particle Physics University of Maryland Elementary Particle Physics
University of Maryland Elementary Particle Physics}
	\end{picture}
	\par\vskip-8mm}
\def\bordero{						
	\setlength{\unitlength}{1mm}
	\newcount\xco
	\newcount\yco
	\xco=-24
	\yco=12
	\begin{picture}(140,0)
	\put(\xco,\yco){$\ktl$}
	\advance\yco by-1
	{\loop
	\put(\xco,\yco){$\kcl$}
	\advance\yco by-2
	\ifnum\yco>-240
	\repeat
	\put(\xco,\yco){$\kbl$}}
	\xco=158
	\yco=12
	\put(\xco,\yco){$\ktr$}
	\advance\yco by-1
	{\loop
	\put(\xco,\yco){$\kcr$}
	\advance\yco by-2
	\ifnum\yco>-240
	\repeat
	\put(\xco,\yco){$\kbr$}}
	\put(-20,12){\ooo bacdefghidfghghdhededbihdgdfdfhhdheidhdhebaaahjhhdahbahgdedgehgfdiehhgdigicba}
	\put(-20,-241.5){\ooo ababaighefdbfghgeahgdfgafagihdidihiidhiagfedhadbfdecdcdfagdcbhaddhbgfchbgfdacfediacbabab}
	\end{picture}
	\par\vskip-8mm}
\def\headpic{						
	\indent
	\setlength{\unitlength}{.4mm}
	\thinlines
	\par
	\begin{picture}(29,16)
	\put(165,16){\line(1,0){4}}
	\put(170,16){\line(1,0){4}}
	\put(180,16){\line(1,0){4}}
	\put(175,0){\line(1,0){4}}
	\put(180,0){\line(1,0){4}}
	\put(185,0){\line(1,0){4}}
	\put(169,0){\line(0,1){16}}
	\put(170,0){\line(0,1){16}}
	\put(179,0){\line(0,1){16}}
	\put(180,0){\line(0,1){16}}
	\put(184,0){\line(0,1){16}}
	\put(185,0){\line(0,1){16}}
	\put(169,16){\oval(8,32)[bl]}
	\put(170,16){\oval(8,32)[br]}
	\put(179,0){\oval(8,32)[tl]}
	\put(185,0){\oval(8,32)[tr]}
	\end{picture}
	\par\vskip-6.5mm
	\thicklines}
\def\title#1#2#3#4{\border\headpic {\hbox to\hsize{#4 \hfill UMDEPP #3}}\par
	\begin{center} \vglue .5in {\large\bf #1}\\[.6in] 
	{#2}\\[.1in] {\it Department of Physics and Astronomy}\\
	{\it University of Maryland, College Park, MD 20742}\\[1.5in] 
	{\bf Abstract}\\[.1in] \end{center} \begin{quotation}}	
\def\Title#1#2#3#4#5#6#7{\border\headpic 
	{\hbox to\hsize{#7 \hfill UMDEPP #6}}\par
	\begin{center} \vglue .4in {\large\bf #1}\\[.4in] 
	{#2}\\[.1in] {\it Department of Physics and Astronomy}\\
	{\it University of Maryland, College Park, MD 20742}\\[.1in]
	{#3}\\[.1in] {\it {#4}}\\ {\it {#5}}\\[.5in] {\bf Abstract}\\[.1in]
	\end{center} \begin{quotation}}			
\def\endtitle{\end{quotation}\newpage}			

\def\sect#1{\bigskip\medskip \goodbreak \noindent{\bf {#1}} \nobreak \medskip}
\def\refs{\sect{References} \footnotesize \frenchspacing \parskip=0pt}
\def\Item{\par\hang\textindent}

\def\[{\lfloor{\hskip 0.35pt}\!\!\!\lceil\,}
\def\]{\,\rfloor{\hskip 0.35pt}\!\!\!\rceil}

\def\Lag{{\cal L}}
\def\du#1#2{_{#1}{}^{#2}}
\def\ud#1#2{^{#1}{}_{#2}}

\def\plpl{{+\!\!\!\!\!{\hskip 0.009in}{\raise -1.0pt\hbox{$_+$}}
{\hskip 0.0008in}}} 
\def\mimi{{-\!\!\!\!\!{\hskip 0.009in}{\raise -1.0pt\hbox{$_-$}}
{\hskip 0.0008in}}}

\def\pl#1#2#3{Phys.~Lett.~{\bf {#1}B} (19{#2}) #3}
\def\np#1#2#3{Nucl.~Phys.~{\bf B{#1}} (19{#2}) #3}
\def\prl#1#2#3{Phys.~Rev.~Lett.~{\bf #1} (19{#2}) #3}
\def\pr#1#2#3{Phys.~Rev.~{\bf D{#1}} (19{#2}) #3}
\def\cqg#1#2#3{Class.~and Quant.~Gr.~{\bf {#1}} (19{#2}) #3}

\def\prep#1#2#3{Phys.~Rep.~{\bf {#1}C} (19{#2}) #3}

\def\ijmp#1#2#3{Int.~Jour.~Mod.~Phys.~{\bf A{#1}} (19{#2}) #3}

\def\ibid#1#2#3{{\it ibid.}~{\bf {#1}} (19{#2}) #3}

\def\mpl#1#2#3{Mod.~Phys.~Lett.~{\bf A{#1}} (19{#2}) #3}

\def\eqques{{~\,={\hskip -11.5pt}\raise -1.8pt\hbox{\large ?}
{\hskip 4.5pt}\,}}

\def\fracmm#1#2{{{#1}\over{#2}}}

\def\half{{\fracm12}}

\def\frac#1#2{{\textstyle{#1\over\vphantom2\smash{\raise -.20ex
	\hbox{$\scriptstyle{#2}$}}}}}			
\def\fracm#1#2{\hbox{\large{${\frac{{#1}}{{#2}}}$}}}

\def\Dot#1{\buildrel{_{_{\hskip 0.01in}\bullet}}\over{#1}}

\def\Tilde#1{{\widetilde{#1}}\hskip 0.015in}	
\def\Hat#1{\widehat{#1}}			

\def\scst{\scriptstyle}

\def\.{.$\,$}

\def\hatm{\hat m} \def\hatn{\hat n} \def\hatr{\hat r}
\def\hats{\hat s}

\def\un{\underline} 
\def\-{{\hskip 1.5pt}\hbox{-}}

\def\kd#1#2{\d\du{#1}{#2}}
\def\footnotew#1{\footnote{\hsize=6.5in {#1}}} 

\def\low#1{{\raise -3pt\hbox{${\hskip 1.0pt}\!_{#1}$}}}

\def\Dot#1{\buildrel{_{_{\hskip 0.01in}\bullet}}\over{#1}}

\begin{document}

\font\tenmib=cmmib10
\font\sevenmib=cmmib10 at 7pt 
\font\fivemib=cmmib10 at 5pt  
\font\tenbsy=cmbsy10
\font\sevenbsy=cmbsy10 at 7pt 
\font\fivebsy=cmbsy10 at 5pt  
\def\BMfont{\textfont0\tenbf \scriptfont0\sevenbf
                              \scriptscriptfont0\fivebf
            \textfont1\tenmib \scriptfont1\sevenmib
                               \scriptscriptfont1\fivemib
            \textfont2\tenbsy \scriptfont2\sevenbsy
                               \scriptscriptfont2\fivebsy}
\def\rlx{\relax\leavevmode}                  
\def\BM#1{\rlx\ifmmode\mathchoice
                      {\hbox{$\BMfont#1$}}
                      {\hbox{$\BMfont#1$}}
                      {\hbox{$\scriptstyle\BMfont#1$}}
                      {\hbox{$\scriptscriptstyle\BMfont#1$}}
                 \else{$\BMfont#1$}\fi}

\font\tenmib=cmmib10
\font\sevenmib=cmmib10 at 7pt 
\font\fivemib=cmmib10 at 5pt  
\font\tenbsy=cmbsy10
\font\sevenbsy=cmbsy10 at 7pt 
\font\fivebsy=cmbsy10 at 5pt  
\def\BMfont{\textfont0\tenbf \scriptfont0\sevenbf
                              \scriptscriptfont0\fivebf
            \textfont1\tenmib \scriptfont1\sevenmib
                               \scriptscriptfont1\fivemib
            \textfont2\tenbsy \scriptfont2\sevenbsy
                               \scriptscriptfont2\fivebsy}
\def\BM#1{\rlx\ifmmode\mathchoice
                      {\hbox{$\BMfont#1$}}
                      {\hbox{$\BMfont#1$}}
                      {\hbox{$\scriptstyle\BMfont#1$}}
                      {\hbox{$\scriptscriptstyle\BMfont#1$}}
                 \else{$\BMfont#1$}\fi}

\def\inbar{\vrule height1.5ex width.4pt depth0pt}
\def\sinbar{\vrule height1ex width.35pt depth0pt}
\def\ssinbar{\vrule height.7ex width.3pt depth0pt}
\font\cmss=cmss10
\font\cmsss=cmss10 at 7pt
\def\ZZ{\rlx\leavevmode
             \ifmmode\mathchoice
                    {\hbox{\cmss Z\kern-.4em Z}}
                    {\hbox{\cmss Z\kern-.4em Z}}
                    {\lower.9pt\hbox{\cmsss Z\kern-.36em Z}}
                    {\lower1.2pt\hbox{\cmsss Z\kern-.36em Z}}
               \else{\cmss Z\kern-.4em Z}\fi}
\def\Ik{\rlx{\rm I\kern-.18em k}}  
\def\IC{\rlx\leavevmode
             \ifmmode\mathchoice
                    {\hbox{\kern.33em\inbar\kern-.3em{\rm C}}}
                    {\hbox{\kern.33em\inbar\kern-.3em{\rm C}}}
                    {\hbox{\kern.28em\sinbar\kern-.25em{\rm C}}}
                    {\hbox{\kern.25em\ssinbar\kern-.22em{\rm C}}}
             \else{\hbox{\kern.3em\inbar\kern-.3em{\rm C}}}\fi}
\def\IP{\rlx{\rm I\kern-.18em P}}
\def\IR{\rlx{\rm I\kern-.18em R}}
\def\IN{\rlx{\rm I\kern-.20em N}}
\def\Ione{\rlx{\rm 1\kern-2.7pt l}}

%
%
\def\unredoffs{} \def\redoffs{\voffset=-.31truein\hoffset=-.59truein}
\def\speclscape{\special{ps: landscape}}

\newbox\leftpage \newdimen\fullhsize \newdimen\hstitle \newdimen\hsbody
\tolerance=1000\hfuzz=2pt\def\fontflag{cm}
\catcode`\@=11 
\doit0
{
\def\bigans{b }
\message{ big or little (b/l)? }\read-1 to\answ
\ifx\answ\bigans\message{(This will come out unreduced.}
}
\hsbody=\hsize \hstitle=\hsize 
\doit0{
\else\message{(This will be reduced.} \let\l@r=L
\redoffs \hstitle=8truein\hsbody=4.75truein\fullhsize=10truein\hsize=\hsbody
\output={\ifnum\pageno=0 
  \shipout\vbox{\speclscape{\hsize\fullhsize\makeheadline}
    \hbox to \fullhsize{\hfill\pagebody\hfill}}\advancepageno
  \else
  \almostshipout{\leftline{\vbox{\pagebody\makefootline}}}\advancepageno
  \fi}
}
\def\almostshipout#1{\if L\l@r \count1=1 \message{[\the\count0.\the\count1]}
      \global\setbox\leftpage=#1 \global\let\l@r=R
 \else \count1=2
  \shipout\vbox{\speclscape{\hsize\fullhsize\makeheadline}
      \hbox to\fullhsize{\box\leftpage\hfil#1}}  \global\let\l@r=L\fi}
\fi
\def\nolabels{\def\wrlabeL##1{}\def\eqlabeL##1{}\def\reflabeL##1{}}
\def\writelabels{\def\wrlabeL##1{\leavevmode\vadjust{\rlap{\smash%
{\line{{\escapechar=` \hfill\rlap{\sevenrm\hskip.03in\string##1}}}}}}}%
\def\eqlabeL##1{{\escapechar-1\rlap{\sevenrm\hskip.05in\string##1}}}%
\def\reflabeL##1{\noexpand\llap{\noexpand\sevenrm\string\string\string##1}}}
\nolabels
%
\global\newcount\secno \global\secno=0
\global\newcount\meqno \global\meqno=1
\def\newsec#1{\global\advance\secno by1\message{(\the\secno. #1)}
\global\subsecno=0\eqnres@t\noindent{\bf\the\secno. #1}
\writetoca{{\secsym} {#1}}\par\nobreak\medskip\nobreak}
\def\eqnres@t{\xdef\secsym{\the\secno.}\global\meqno=1\bigbreak\bigskip}
\def\sequentialequations{\def\eqnres@t{\bigbreak}}\xdef\secsym{}
\global\newcount\subsecno \global\subsecno=0
\def\subsec#1{\global\advance\subsecno by1\message{(\secsym\the\subsecno. #1)}
\ifnum\lastpenalty>9000\else\bigbreak\fi
\noindent{\it\secsym\the\subsecno. #1}\writetoca{\string\quad
{\secsym\the\subsecno.} {#1}}\par\nobreak\medskip\nobreak}
\def\appendix#1#2{\global\meqno=1\global\subsecno=0\xdef\secsym{\hbox{#1.}}
\bigbreak\bigskip\noindent{\bf Appendix #1. #2}\message{(#1. #2)}
\writetoca{Appendix {#1.} {#2}}\par\nobreak\medskip\nobreak}
%
%
\def\eqnn#1{\xdef #1{(\secsym\the\meqno)}\writedef{#1\leftbracket#1}%
\global\advance\meqno by1\wrlabeL#1}
\def\eqna#1{\xdef #1##1{\hbox{$(\secsym\the\meqno##1)$}}
\writedef{#1\numbersign1\leftbracket#1{\numbersign1}}%
\global\advance\meqno by1\wrlabeL{#1$\{\}$}}
\def\eqn#1#2{\xdef #1{(\secsym\the\meqno)}\writedef{#1\leftbracket#1}%
\global\advance\meqno by1$$#2\eqno#1\eqlabeL#1$$}
%
\newskip\footskip\footskip14pt plus 1pt minus 1pt 
\def\footnotefont{\ninepoint}\def\f@t#1{\footnotefont #1\@foot}
\def\f@@t{\baselineskip\footskip\bgroup\footnotefont\aftergroup\@foot\let\next}
\setbox\strutbox=\hbox{\vrule height9.5pt depth4.5pt width0pt}
\global\newcount\ftno \global\ftno=0
\def\foot{\global\advance\ftno by1\footnote{$^{\the\ftno}$}}
%
\newwrite\ftfile
\def\footend{\def\foot{\global\advance\ftno by1\chardef\wfile=\ftfile
$^{\the\ftno}$\ifnum\ftno=1\immediate\openout\ftfile=foots.tmp\fi%
\immediate\write\ftfile{\noexpand\smallskip%
\noexpand\item{f\the\ftno:\ }\pctsign}\findarg}%
\def\footatend{\vfill\eject\immediate\closeout\ftfile{\parindent=20pt
\centerline{\bf Footnotes}\nobreak\bigskip\input foots.tmp }}}
\def\footatend{}
%
%
\global\newcount\refno \global\refno=1
\newwrite\rfile
%
\def\ref{[\the\refno]\nref}%
\def\nref#1{\xdef#1{[\the\refno]}\writedef{#1\leftbracket#1}%
\ifnum\refno=1\immediate\openout\rfile=refs.tmp\fi%
\global\advance\refno by1\chardef\wfile=\rfile\immediate%
\write\rfile{\noexpand\Item{#1}\reflabeL{#1\hskip.31in}\pctsign}\findarg\hskip10.0pt}%
\def\findarg#1#{\begingroup\obeylines\newlinechar=`\^^M\pass@rg}
{\obeylines\gdef\pass@rg#1{\writ@line\relax #1^^M\hbox{}^^M}%
\gdef\writ@line#1^^M{\expandafter\toks0\expandafter{\striprel@x #1}%
\edef\next{\the\toks0}\ifx\next\em@rk\let\next=\endgroup\else\ifx\next\empty%
\else\immediate\write\wfile{\the\toks0}\fi\let\next=\writ@line\fi\next\relax}}
\def\striprel@x#1{} \def\em@rk{\hbox{}}
\def\lref{\begingroup\obeylines\lr@f}
\def\lr@f#1#2{\gdef#1{\ref#1{#2}}\endgroup\unskip}
\def\semi{;\hfil\break}
\def\addref#1{\immediate\write\rfile{\noexpand\item{}#1}} 
\def\listrefs{\footatend\vfill\supereject\immediate\closeout\rfile\writestoppt
\baselineskip=14pt\centerline{{\bf References}}\bigskip{\frenchspacing%
\parindent=20pt\escapechar=` \input refs.tmp\vfill\eject}\nonfrenchspacing}
%
\def\listrefsr{\immediate\closeout\rfile\writestoppt
\baselineskip=14pt\centerline{{\bf References}}\bigskip{\frenchspacing%
\parindent=20pt\escapechar=` \input refs.tmp\vfill\eject}\nonfrenchspacing}
\def\startrefs#1{\immediate\openout\rfile=refs.tmp\refno=#1}
\def\xref{\expandafter\xr@f}\def\xr@f[#1]{#1}
\def\refs#1{\count255=1[\r@fs #1{\hbox{}}]}
\def\r@fs#1{\ifx\und@fined#1\message{reflabel \string#1 is undefined.}%
\nref#1{need to supply reference \string#1.}\fi%
\vphantom{\hphantom{#1}}\edef\next{#1}\ifx\next\em@rk\def\next{}%
\else\ifx\next#1\ifodd\count255\relax\xref#1\count255=0\fi%
\else#1\count255=1\fi\let\next=\r@fs\fi\next}
\def\figures{\centerline{{\bf Figure Captions}}\medskip\parindent=40pt%
\def\fig##1##2{\medskip\item{Fig.~##1.  }##2}}
%
\newwrite\ffile\global\newcount\figno \global\figno=1
\def\fig{fig.~\the\figno\nfig}
\def\nfig#1{\xdef#1{fig.~\the\figno}%
\writedef{#1\leftbracket fig.\noexpand~\the\figno}%
\ifnum\figno=1\immediate\openout\ffile=figs.tmp\fi\chardef\wfile=\ffile%
\immediate\write\ffile{\noexpand\medskip\noexpand\item{Fig.\ \the\figno. }
\reflabeL{#1\hskip.55in}\pctsign}\global\advance\figno by1\findarg}
\def\listfigs{\vfill\eject\immediate\closeout\ffile{\parindent40pt
\baselineskip14pt\centerline{{\bf Figure Captions}}\nobreak\medskip
\escapechar=` \input figs.tmp\vfill\eject}}
\def\xfig{\expandafter\xf@g}\def\xf@g fig.\penalty\@M\ {}
\def\figs#1{figs.~\f@gs #1{\hbox{}}}
\def\f@gs#1{\edef\next{#1}\ifx\next\em@rk\def\next{}\else
\ifx\next#1\xfig #1\else#1\fi\let\next=\f@gs\fi\next}
\newwrite\lfile
{\escapechar-1\xdef\pctsign{\string\%}\xdef\leftbracket{\string\{}
\xdef\rightbracket{\string\}}\xdef\numbersign{\string\#}}
\def\writedefs{\immediate\openout\lfile=labeldefs.tmp \def\writedef##1{%
\immediate\write\lfile{\string\def\string##1\rightbracket}}}
\def\writestop{\def\writestoppt{\immediate\write\lfile{\string\pageno%
\the\pageno\string\startrefs\leftbracket\the\refno\rightbracket%
\string\def\string\secsym\leftbracket\secsym\rightbracket%
\string\secno\the\secno\string\meqno\the\meqno}\immediate\closeout\lfile}}
\def\writestoppt{}\def\writedef#1{}
\def\seclab#1{\xdef #1{\the\secno}\writedef{#1\leftbracket#1}\wrlabeL{#1=#1}}
\def\subseclab#1{\xdef #1{\secsym\the\subsecno}%
\writedef{#1\leftbracket#1}\wrlabeL{#1=#1}}
\newwrite\tfile \def\writetoca#1{}
\def\leaderfill{\leaders\hbox to 1em{\hss.\hss}\hfill}
\def\writetoc{\immediate\openout\tfile=toc.tmp
   \def\writetoca##1{{\edef\next{\write\tfile{\noindent ##1
   \string\leaderfill {\noexpand\number\pageno} \par}}\next}}}
\def\listtoc{\centerline{\bf Contents}\nobreak\medskip{\baselineskip=12pt
 \parskip=0pt\catcode`\@=11 \input toc.tex \catcode`\@=12 \bigbreak\bigskip}}
\catcode`\@=12 
%

\font\fifteenrm=cmr12 at 14.0pt
\font\fifteenrm=cmr10 at 15.0pt

\def\alephnull{~$\large{\aleph_0}\,$~} 
\def\spinorip#1{\left({#1}\right)} 
\def\Check#1{\raise0.6pt\hbox{\Large\v{}}{\hskip -8.5pt}{#1}}
\def\kd#1#2{\d\du{#1}{#2}}

\def\fracm#1#2{\,\hbox{\large{${\frac{{#1}}{{#2}}}$}}\,}
\def\fracmm#1#2{\,{{#1}\over{#2}}\,}

\def\framing#1{\doit{#1}
{\framingfonts{#1}
\border\headpic 
}}

\framing{0}

~~~

\vskip 0.07in

{\hbox to\hsize{
June, 1996
\hfill UMDEPP 96--109}}
{\hbox to\hsize{~~~~~ ~~~~~~\hfill 
}} 
\par

\hsize=6.5in
\textwidth=6.5in

\begin{center}
\vglue 0.1in

{\large\bf ${\aleph_0}\,$-$\,$Extended ~Supergravity ~and 
~Chern$\,$-$\,$Simons ~Theories}$\,$\footnote{This work is supported in part
by NSF grant \# PHY-93-41926, 
and by DOE grant \# DE-FG02-94ER40854  
}  \\[.1in]

\baselineskip 10pt 

\vskip 0.18in

\doit1{Hitoshi ~N{\small ISHINO}\footnotew{E-mail: nishino@umdhep.umd.edu}
~ and~ S.~James ~G{\small ATES}, $\,$Jr.\footnote{E-mail:
gates@umdhep.umd.edu}   \\[.25in]
{\it Department of Physics} \\ [.015in]
{\it University of Maryland at College Park}\\ [.015in]
{\it College Park, MD 20742-4111, USA} \\[.18in]
}

\vskip 0.8in

{\bf Abstract} \\[0.1in]  
\end{center}

\begin{quotation}

~~~We give generalizations of extended Poincar\'e supergravity with 
{\it arbitrarily many} supersymmetries in the absence of central charges 
in three-dimensions by gauging its intrinsic global $~SO(N)$~ symmetry.  
We call these \alephnull (Aleph-Null) supergravity theories.  We further 
couple a non-Abelian supersymmetric Chern-Simons theory and an Abelian 
topological BF theory to \alephnull supergravity.  Our result overcomes 
the previous difficulty for supersymmetrization of Chern-Simons theories 
beyond $~N=4$.  This feature is peculiar to the Chern-Simons and BF 
theories including supergravity in three-dimensions.  We also show that 
dimensional reduction schemes for four-dimensional theories such as 
$~N=1$~ self-dual supersymmetric Yang-Mills theory or $~N=1$~ supergravity 
theory that can generate \alephnull globally and locally supersymmetric 
theories in three-dimensions.  As an interesting application, we present 
\alephnull supergravity Liouville theory in two-dimensions 
after appropriate dimensional reduction from three-dimensions.      
   
\endtitle

\oddsidemargin=0.03in
\evensidemargin=0.01in
\hsize=6.5in
\textwidth=6.5in
\baselineskip 18.0pt

\centerline{\bf 1.~~Introduction}

Recently there have been new developments in globally supersymmetric 
theories in three dimensions (3D) or lower dimensions \ref\gr{S.J.~Gates, 
Jr.~and L.~Rana, \pl{352}{95}{50}; \ibid{369B}{96}{269}.} based 
on ${\cal G}{\cal R}
({\rm d}, N)$ algebras leading to the use of quantities called $~L$~ and 
$~R$~ matrices satisfying a certain anti-commutator algebra which 
generalizes the usual Clifford algebra.  Representations of these
algebras enable us to construct a {\it {theory}} of on-shell 3D
representations as well as off-shell 1D representations. In particular
scalar multiplets with arbitrarily large numbers of supersymmetries have 
been constructed \ref\dew{B. de~Wit, A.K.~Tollst\' en and H.~Nicolai,
\np{363}{91}{221}.}.  We can call these systems \alephnull 
(alephnull) supersymmetry, since in the limit $~N\rightarrow\infty$
they can accommodate infinitely many supersymmetries.  An interesting question 
then is whether there is a similar technique applicable to Chern-Simons (CS) 
theories.  If the answer is affirmative, then the subsequent question is 
whether those globally supersymmetric theories can be coupled to supergravity 
with \alephnull supersymmetry.  

As a matter of fact, there has been indication that supergravity theories
in 3D can be interpreted as CS theories, in particular, with
infinitely many extended supersymmetries \ref\alephnullorig{P.~van 
Nieuwenhuizen, \pr{32}{85}872; A.~Ach\'ucarro
and P.K.~Townsend, \pl{180}{86}{89}; E.~Witten, \np{311}{88}{46};
J.~Horne and E.~Witten, \prl{62}{89}{501}.}.  In a paper in a similar direction 
P.~Howe {\it et. al.}~\ref\howe{P.S.~Howe, J.M.~Izquierdo, G.~Papadopoulos 
and P.K.~Townsend,
King's College - DAMTP preprint, R/95/13 (May 1995).} it is found that there
exist infinitely many local supersymmetries for on-shell Poincar\'e
supergravity, with two sets of vector fields, one set gauging the group 
$~SO(p)\otimes SO(q)$~ with $~N=p+q$~ and another gauging the central 
charges \howe.  This system is analogous to the conformal supergravity with 
arbitrary number of extended supersymmetries \ref\rn{M.~Ro{\v c}ek and
P.~van Nieuwenhuizen, \cqg{3}{86}{43}.}\ref\ng{H.~Nishino and S.J.~Gates,
\ijmp{8}{93}{3371}.}.                

Independent of these developments within 3D, there has been
another important observation \ref\wittendual{E.~Witten, \mpl{10}{95}{2153}.} 
about ``strong-weak coupling duality'' 
between 3D superstring and 4D superstring theories in order to
understand the vanishing of the cosmological constant in 4D.  
According to this scenario, the reason we have exactly zero cosmological
constant in 4D even after supersymmetry breaking is due to the duality 
between these 4D theories and 3D superstring theories in which 
supersymmetry keeps the zero cosmological constant, while the usual 
mass degeneracy between bosons and fermions is lifted.     

Considering these recent developments, it is 
crucial to find supersymmetric non-Abelian CS
and topological BF theories \ref\rakowski{{\it See, e.g.,} D.~Birmingham,
M.~Blau, M.~Rakowski and G.~Thompson, \prep{209}{91}{129}.} that can 
couple to the \alephnull Poincar\'e supergravity.  We will try to
combine the two different recent theories, {\it i.e.}, one with arbitrary number
of global supersymmetries in terms of $~L$~ and $~R$~ matrices \gr, and
supergravity theories based on CS formulations \alephnullorig.  

In our previous paper \ng\ we found an apparent barrier that prevented
going beyond $~N=4$~ supersymmetric non-Abelian CS theories.  In the present 
paper, we will present two ways to bypass this difficulty, maintaining 
the on-shell closure of the gauge algebra by virtue of vanishing field 
strengths of the gravitini.  We will see the minimal field content needed 
for on-shell closure of Poincar\'e supergravity with no central charges 
\howe.  After understanding this extended supergravity, we also
perform its coupling to supersymmetric CS theory, a topological BF theory
\rakowski, and also to a tensor multiplet with arbitrary number of
supersymmetries.  As by-products and interesting applications, we also 
present \alephnull supergravity Liouville theory in 2D, as well as \alephnull BF
theory in 3D.

\bigskip\bigskip\bigskip


\centerline{\bf 2.~~ \alephnull Supergravity in 3D}

We start with reviewing the on-shell \alephnull extended supergravity in 
3D \alephnullorig.   This supergravity multiplet consists of the 
dreibein and the gravitini $~(e\du\m m,\psi\du\m A)$, where the indices
$~{\scst \m,~\n,~ \cdots ~=~ 0,~ \cdots,~3}$~ are the curved coordinates,
while  $~{\scst m,~n,~ \cdots ~=~ (0),~ \cdots,~(3)}$~ are local Lorentz
coordinates.  Relevantly the signature of our space-time is $~(+,-,-)$,
while our $~\g\-$matrices satisfy $~\g^{m n
r} = i \e^{m n r}$~ with $~\e^{(0)(1)(2)} = +1$, and 
$~\g^m \g^n + \g^n \g^m = 2 \eta^{m n}$.      
We use the indices $~{\scst
A,~B,~\cdots~=~1,~\cdots,~N}$~ for the $~N\-$extended supersymmetries.   
Since our formulation is on-shell, we do not have any additional gauge fields 
such as $~A^{A B}$~ or $~C^{A B}$~ presented in the off-shell formulation in
ref.~\howe.     

The supersymmetry transformation rules and the invariant lagrangian
\alephnullorig\ are similar to the most standard form of $~N=1$~
supergravity in 4D \ref\ffvn{D.~Freedman, S.~Ferrara and P.~van
Nieuwenhuizen, \pr{14}{76}{912}.} or 3D \howe\ref\ms{N.~Marcus and
J.H.~Schwarz, \np{228}{83}{145}.}:    
$$\eqalign{&\d_Q e\du\m m = - i (\Bar\e{}^A \g^m \psi\du\m A) ~~, \cr 
&\d_Q\psi\du\m A = \partial_\m \e^A + \fracm14 \Hat \omega\du\m{m n}(e,\psi)
\g_{m n} \e^A \equiv D_\m(\Hat \omega) \e^A~~, \cr } 
\eqno(2.1) $$
where $~\omega\du\m{r s}$~ has the $~\psi\-$torsion like 4D \ref
\pvn{{\it See e.g.}, P.~van Nieuwenhuizen, \prep{68}{81}{189}.}:
$$ \Hat\omega\du \m {r s } (e,\psi) = \half \left( C\du\m {r s} -
C\du\m{s r} + C\ud{s r}\m \right)  ~~, ~~~~
C\du{\m\n} m \equiv \partial_\m e\du\n m - \partial_\n e\du \m m + i\left( 
\Bar\psi_\m{}^A \g^m \psi _\n{}^A \right) ~~. 
\eqno(2.2) $$
with the invariant lagrangian:
$$\Lag _{\aleph_0{\rm S G}} = - \fracm14 e R(\Hat \omega) -\fracm 1 4 
\e^{\m\n\r}\left(\Bar\psi_\m{}^A {\cal R}\du{\n\r}A \right)  ~~,   
\eqno(2.3) $$   
where $~{\cal R}\du{\m\n}A $~ is the gravitino field strength:
$${\cal R}_{\m\n}{}^A \equiv D_\m (\Hat\omega) \psi\du\n A-
D_\n (\Hat\omega) \psi\du\m A~~
\eqno(2.4) $$

Since this system has been presented in the past \alephnullorig\ we do not 
repeat the details.  Such infinitely many supersymmetries are possible due
to the peculiar feature of 3D namely both the dreibein and gravitini
have no physical degrees of freedom.  The closure of two
supersymmetries on the dreibein is the usual one, while that on the 
gravitino yields a peculiar ``extra'' transformation on $~\psi\du\m A$, {\it
i.e.}, $~\[\,\d_Q (\e_1), \d_Q(\e_2) \,\] =\d_P (i \Bar\e_1{}^A\g^m\e_2{}^A)
+  \d_{\rm E}$~ with    
$$\eqalign{\d_{\rm E} \psi_\m{}^A = \, & + i e^{-1} \e\du\m{\n\r} 
A\du1{A B} \g_\n \Tilde {\cal R}\du\r B + \fracm1{12} A\du 1{A B} \Tilde 
{\cal R}\du\m B\cr   
& -\fracm38 e^{-1} \e\du{\m\n} \r S\du1{\n \,A B} \Tilde {\cal R}\du\r B
+\fracm18  e^{-1} \e\du{\m\n}\r S\du1\n \Tilde {\cal R}\du\r A \cr 
& + S\du {2\m}{A B} (i\g_\n \Tilde{\cal R}^{\n B} ) + S\du2{\n A B} \left(i
\g_\m\Tilde {\cal R}\du\n B - i\g_\n\Tilde {\cal R} \du\m B \right)
~~,\cr}   
\eqno(2.5) $$
where 
$$\li{&\Tilde{\cal R}{}^{\m A} \equiv e^{-1} \e^{\m\n\r} {\cal
R}_{\n\r}{}^A~~, \cr    
& A\du1{A B} \equiv \fracm 3 8 \left(\Bar\e\du1 {\[ A} \e\du2
{B\]} \right)  = - A \du1{A B} ~~,  ~~~~ 
S\du0{A B} \equiv - i\left( \Bar\e_1^{(A} \g^\m \e_2^{B)}
\right) = + S\du 0 {B A} ~~. 
&(2.6)  \cr }  $$ 
The lagrangian (2.2) is invariant under these extra symmetries (2.5), as is
easily confirmed.  This is natural because these extra 
symmetries can be regarded as the {\it on-shell} vanishing terms in the
commutator algebra  \pvn, since the gravitino field equations are simply
$~{\cal R}_{\m\n} = 0$~ anyway.       

Compared also with the algebra
presented in \howe, our system lacks vector fields due to the absence of 
central charges gauged by $~C_\m{}^{A B}$ in the former.  Since the two sets of
vector fields, {\it i.e.}, one for $~SO(p)\otimes SO(q)$~ and another for the
central charges \howe, appear in pair in the lagrangian, it is natural that 
our system does not have any of theses vector fields.  In any case, 
our system has the minimal field content for the Poincar\'e algebra with no
central charges.       

We now present the following generalizations of the ``minimal'' \alephnull
supergravity above, which have not been given in the past to our knowledge.   
First we can gauge the global $~SO(N)$~ symmetry by the
gauge field $~B\du\m {A B}$~ and an additional vector field $~C\du\m{A B}$.  
The resulting multiplet $~(e\du\m m, \psi\du\m A, B\du\m {A B}, 
C\du\m{A B})$~ can be obtained from the supergravity multiplet in \howe\ 
by identifying their central charges $~Z^{i j}$~ identified with the $~SO(N)$~
generators  $~T^{i j}$.  We first include the $~SO(N)$~ minimal coupling in
all the derivatives such as   $$ 2{\cal D}_{\[\m} \psi\du{\n\]} A \equiv 2
\left( D_{\[\m} (\Hat\omega) \psi\du{\n\]} A + \Tilde g B\du{\[\m} {A B} 
\psi \du{\n\]} B \right) \equiv {\Check R} \du{\m\n} A ~~,  
\eqno(2.7) $$
and the similarly for $~D(\Hat\omega)_\m \e^A $~ in (2.1) by $~{\cal D}_\m
\e^A$.  We also add the transformations 
$$\eqalign{& \d_Q B\du\m{A B} = \fracm i 2 \left( {\Bar\e}{}^{\[A |} \g^\n
\Check{\cal R}\du{\m\n} {|B\]} \right) + \half e^{-1} \e\du \m{\r\s} \left(
{\Bar\e}{} ^{\[ A} \Check{\cal R} \du{\r\s}{B \]} \right) ~~, \cr
& \d_Q C\du\m {A B} = - \left({\Bar\e}{}^{\[ A} \psi \du \m {B\]} \right) 
~~, \cr}  
\eqno(2.8) $$   
together with the new additional terms in the gravitino transformation:
$$\d_Q\psi\du\m A = \partial_\m \e^A + \fracm1 4 \Hat\omega\du
\m{m n} \g_{m n} \e^A + \Tilde g A\du\m {A B} \e^B  
- \Tilde g e^{-1} \e\du\m {\r\s} \Hat H\du{\r\s} {A B}  + i \Tilde g \g_\n
\e^B \Hat H\du{\m\n}{A B} ~~,  
\eqno(2.9) $$  
where 
$$ H \du{\m\n} {A B} \equiv \left( \partial_\m C\du\n{A B} + 2 B\du\m{\[
A|C} C\du\n{C|B\]} \right) - {\scst (\m \leftrightarrow \n )} ~~,
\eqno(2.10) $$
looks like a field strength and is actually covariant under the $~SO(N)$, 
but different from the proper $~SO(N)$~ gauge field strength:
$$ G \du{\m\n} {A B} = \left( \partial_\m B \du\n{A B} + B\du\m{A C} B\du
\n{C B} \right) - {\scst (\m \leftrightarrow \n )} ~~.  
\eqno(2.11) $$
As usual, the {\it hatted} quantities are supercovariant, {\it e.g.},
$$ \Hat H\du{\m\n} {A B} \equiv H\du{\m\n} {A B} + \half 
\left( \Bar\psi{}\du\m{\[ A} \psi\du\n{B \]} \right) ~~. 
\eqno(2.12) $$  
  
Finally our lagrangian has  
an explicit $~\Tilde g\-$term like a $~B F$~ lagrangian \rakowski\howe:
$$ \Lag _{\aleph_0{\rm SG},\,\Tilde g} = - \fracm1 4 e R(\Hat \omega) 
- \fracm1 4
\e^{\m\n\r} \left(\Bar\psi\du\m A {\Check{\cal R}}\du{\r\s} A \right) + 
\Tilde g \e^{\m\n\r} C\du\m{A B} G\du{\r\s} {A B} ~~,  
\eqno(2.13) $$
and relevantly $~G\du{\r\s}{A B}$~  
is the ``field strength'' of $~C\du\m{A B}$.  

The on-shell closure of this multiplet is easy to confirm:\footnotew{Any 
``extra'' transformation involved is skipped here.}  
$$ \[  \d_Q (\e_1), \d_Q(\e_2) \] = \d_P (i\Bar\e_1^A \g^m \e_2^A) + 
\d_{\rm G} (\Bar\e_1^{\[A} \e_2^{B\]} ) ~~,  
\eqno(2.14) $$ 
where $~\d_Q$~ is the $~O(N)$~
gauge transformation acting as  
$$ \d_{\rm G} A\du \m{A B} = {\cal D}_\m 
\L^{A B}~~, ~~~~ \d_{\rm G} C\du\m{A B} = {\cal D}_\m \L^{A B} ~~ .   
\eqno(2.15)  $$
Note that $~{\cal D}_\m $~ is $~SO(N)$~ covariant with the
minimal coupling by $~A\du\m{A B}$.  Even though both of these fields 
transform in the same way under  $~SO(N)$, there will be no problem in 
3D for the same reason given in the context of $~N=4$~ CS theory in
\ng.  To put it differently, we can identify the central charges with the
$~SO(N)$~ generators, when there is only one ~$SO(N)$~ symmetry.  
Needless to say, we can always go back to the {\it minimal} supergravity 
field content by turning off the $~SO(N)$~ coupling: $~\Tilde g \rightarrow
0$.    
         
There is another generalization which has not been given in
literature.  We can include additional vector and a spinor
fields $~A\du\m I$~ and $~\l$~ with a supersymmetric CS form.  Now 
the new field content is $~(e\du\m m, \psi\du\m A, A\du\m{A B}, B\du\m{A B},
C\du\m{A B}, \l^A)$, where $~B\du\m{A B}$~ is the gauge field for gauging
$~SO(N)$.  We use the indices $~{\scst A,~B,~\cdots~=~1,~\cdots,~N}$~ for 
the vectorial representation of $~SO(N)$.     

Our lagrangian   
$$\eqalign{\Lag_{\aleph_0{\rm SG},\,\Tilde g,\,\Tilde h} = \, & - \fracm 14 e
R(\Hat\omega) + \fracm1 4 \e^{\m\n\r} \left( \Bar\psi{}\du\m A {\cal
R}\du{\n\r} A \right) + \Tilde g \e^{\m\n\r} C\du\m{A B} G\du{\n\r} {A B}\cr
& + \half \Tilde g \Tilde h \left( \Bar\l{}^A\l^A \right) + \half \Tilde g
\Tilde h \e^{\m\n\r} \left( F\du{\m\n} {A B} A\du\r{A B} - \fracm2 3 
A\du\m{A B} A\du\n{B C} A\du\r{C A} \right) ~~, \cr}
\eqno(2.16) $$  
is invariant under the supertranslation rule for this multiplet 
$$\li{&\d_Q e\du\m m= - i \spinorip{\Bar\e{}^A \g^m \psi\du\m A} ~~, \cr 
& \d_Q \psi\du\m A = D_\m(\Hat\omega) \e^A + \Tilde g B\du\m{A B} \e^B 
+ \Tilde g
e^{-1} \e\du\m{\r\s} \e^B \Hat H\du{\r\s} {A B} + i \Tilde g \g^\n \e^B \Hat
H \du{\m\n}{A B} ~~, \cr 
& \d_Q B\du\m{A B} = \fracm i 2 \spinorip{ \Bar\e{}^{\[A|} \g^\n {\cal R}\du
{\m\n}{|B\]} } + \half e^{-1} \e\du\m{\r\s} \spinorip{\Bar\e{}^{\[A} {\cal
R}\du{\r\s} {B\]}}  + i \Tilde h \spinorip{ \Bar\e{}^{\[A} \g_\m \l^{B\]} }
~~, \cr 
& \d_Q C\du\m{A B} = + \half \spinorip{ \Bar\e{}^{\[A} \psi\du\m {B\]} } + i
\Tilde h \spinorip{ \Bar\e{}^{\[A} \g_\m\l^{B\]} } ~~,  
&(2.17) \cr 
& \d_Q A\du \m{A B} = i \spinorip{ \Bar\e{}^{\[A} \g_\m\l^{B\]}  } ~~, \cr 
& \d_Q \l^A = - \g^{\m\n} \e^B \spinorip{ F\du{\m\n}{A B} + G\du{\m\n}{A B}
+ H \du{\m\n}{A B} } 
+ \fracm i 2 \Tilde g \spinorip{ \Bar\e{}^B
\g^\m\psi\du\m B }\l^A + \fracm i 2 \spinorip{ \Bar\e{}^{\[A} {\cal R}
\du{\m\n}{B\]} } (\g_\m\psi\du\n B) ~~. \cr } $$
The $~\Tilde g~$ 	and $~\Tilde h~$ are coupling constants, and in
particular the former is the $~SO(N)$~ coupling.  Therefore if we switch
off $~\Tilde g \rightarrow 0$, then the system is reduced to the {\it
minimal} \alephnull supergravity.  If we keep non-zero $~\tilde g$, while 
taking the limit $~\Tilde h \rightarrow 0$, the two
fields $~A\du\m{A B}$~ and $~\l^A$~ will be removed.  
Our system is thus a combination of the usual supersymmetric CS action
made of the $~A$~ and $~\l\-$fields and the $~SO(N)$~ {\it gauged} 
\alephnull supergravity.           

The relevant (super)field strengths are defined by  
$$\li{&{\cal R} \du{\m\n}
A \equiv\left( \partial_\m\psi_\n{}^A + \fracm1 4 \Hat\omega\du\m{m n} 
\g_{m n} \psi\du\n A + \Tilde g B\du\m{A B} \psi\du\n B \right)  - {\scst
(\m\leftrightarrow \n)}  \cr 
& ~~~~~~ \equiv \left( D_\m \psi\du\n A + i \Tilde g B\du\m {A B} \psi\du
\n B \right) - {\scst (\m \leftrightarrow \n)} \equiv {\cal D}_\m\psi\du\n
A -  {\cal D}_\n\psi\du\m A ~~,  
&(2.18) \cr   
& F\du{\m\n}{A B} \equiv \left(\partial_\m A\du\n{A B} + A\du\m {A C} A
\du\n{C B} \right) - {\scst (\m \leftrightarrow \n)}  ~~, 
&(2.19) \cr 
& G\du{\m\n}{A B} \equiv \left( \partial_\m B\du\n{A B} + \Tilde g B\du
\m{A C} B\du\n{C B} \right) - {\scst (\m \leftrightarrow \n)} ~~, 
&(2.20) \cr 
& H\du{\m\n}{A B} \equiv \left( \partial_\m C\du\n{A B} + \Tilde g B
\du\m{A C} C\du\n{C B} + \Tilde g B\du\m {B C} C\du\n{A C}\right) - {\scst
(\m \leftrightarrow \n)}  ~~, 
&(2.21) \cr  
&\Hat F\du{\m\n}{A B} \equiv F\du{\m\n}{A B} - 2 i \left( \Bar\psi\du{\[\m}
{\[A} \g_{\n\]} \l^{B\]}  \right)   ~~,  
&(2.22) \cr   
&\Hat G\du{\m\n}{A B} \equiv G\du{\m\n}{A B} - i
\spinorip{\Bar\psi\du{\[\m}{\[ A|}\g^\r{\cal R}\du{\n\]\r}{|B\]} } +
e^{-1} \e \du{\[\m} {\r\s} \left(  \Bar\psi\du{\n\]}{\[A} {\cal
R}\du{\r\s}{B\]}\right)  \cr  & ~~~~~ ~~~~~ ~~  - 2i \Tilde h \left(
\Bar\psi\du{\[\m} {\[ A} \g_{\n\]} \l^{B\]} \right) ~~, 
&(2.23) \cr   
&\Hat H\du{\m\n}{A B} \equiv H\du{\m\n}{A B} - \half \left( \bar\psi
\du{\[\m} {\[ A} \psi\du{\n\]}{B\]} \right)  - 2 i \Tilde h \left(
\Bar\psi\du{\[\m}{\[A } \g_{\n\]} \l^{B\]} \right) ~~.  
&(2.24) \cr } $$

\bigskip\bigskip\bigskip

\centerline{\bf 3.~~\alephnull SCS Theory Coupled to \alephnull SG}

Once the \alephnull supergravity is realized, the
next interesting question is its couplings to  any ``matter'' multiplet. 
The easiest case is the CS theory, which has the  simplest lagrangians in
general.  To this end, we have to establish a vector multiplet with
arbitrary number of supersymmetries.  This can be easily done,  once we
notice the duality transformation connecting a scalar multiplet to  a
possible vector multiplet, because in 3D a vector is dual to a
scalar.   As a matter of fact, using the result in \gr, we can establish our
\alephnull non-Abelian vector multiplet $~(A\du{\m\, i} I, \l\du i I)$~ 
coupled to \alephnull supergravity:      $$\eqalign{& \d_Q A\du{\m \, i }I =
+ \fracm1{2{\sqrt2}} \sum_j (L^A)_{i j} \left(\Bar\e{}^A \g_\m \l\du j I
\right) ~~, \cr  &\d_Q \l\du i I = - \fracm1{2{\sqrt2}} \sum_j (R^A)_{i j} (
\g^{\m\n} \e^A) \Hat F\du{\m\n\, j}I  ~~, \cr }
\eqno(3.1) $$   
where ~${\scst I,~J,~\cdots}$~ are for the adjoint representation for
the non-Abelian gauge group, while $~{\scst i,~j,~\cdots~=~1,~2,~\cdots,~d}$~ 
are for the representation of the $~d\times d$~ matrices $~L$~ and $~R$, 
which satisfy the relationships \gr below. These are the defining
conditions for these matrices, 
$$\sum_k \left[ \, (L^A)_{i k} (R^B)_{k j} + (L^B)_{i k} (R^A)_{k j}
\, \right] = - 2 \d^{A B} \d_{i j} ~~, $$  
$$(L^A)_{i j} = - (R^A)_{j i} ~~. 
\eqno(3.2) $$  
The contraction with respect the $~{\scst i,~j,~\cdots}$~ indices always need 
the explicit summation symbols such as $~\sum_i$~ for the reason to be 
seen later.  As has been pointed out in ref.~\gr, we can always construct 
these $~L$~ and $~R$~ matrices for arbitrary $~N$, by choosing a 
sufficiently large $~d\-$dimensional representation.  In particular, 
when $~N=8$~ or $~6~(\hbox{\it mod.}~8)$, these matrices coincide with the 
Clifford algebra construction given in ref.~\ng.  

The field strength of the vector field is defined by 
$$ F\du{\m\n\, i}I \equiv \partial_\m A\du{\n\,i}I - \partial_\n A\du{\m\,i}I 
+ f^{I J K} A\du{\m\,i}J A\du{\n\,i}K ~~, 
\eqno(3.3) $$ 
where $~f^{I J K}$~ is the structure constant of the non-Abelian gauge group.
Due to the third term here with the index $~{\scst i}$~ repeated two
times, we need always the explicit summation symbol for these indices 
to avoid confusion.  In other words, the $i$-index appearing in (3.3) 
should be regarded as {\it {not}} obeying the Einstein-summation
convention. As usual, the {\it hatted} field strength $~\Hat
F\du{\m\n\,i}I$~ denotes its supercovariantization:
$$ \Hat F\du{\m\n\, i} I \equiv F\du{\m\n\, i} I + \bigg[ \,  
\fracm1{2{\sqrt2}} \sum_j (L^A)_{i j} (\Bar\psi_\n{}^A \g_\m \l\du j I)
-{\scst (\m \leftrightarrow \n)} \, \bigg]  ~~.  
\eqno(3.4) $$        
Even though we have $~d$~ multiple gauge fields for a single gauge group, 
this will not  pose any problem.  As a matter of fact, we have already
encountered an exactly the same structure for the case of $~N=4$~ CS
theory in ref.~\ng.  

The gravitino-dependent term in (3.1) is the effect of local
supersymmetry, which does not pose any problem about the closure of the gauge
algebra, as will be seen shortly.  The gravitino-independent terms can be
easily obtained, based on the knowledge about the case of scalar
multiplet with the global \alephnull supersymmetries.    

The invariant lagrangian for our CS theory with 
\alephnull supersymmetries is  
$$\Lag_{\aleph_0\rm C S} =  \half m \e^{\m\n\r} \sum_i \bigg(
F\du{\m\n\,i} I A\du{\r\, i}I - \fracm 1 3 f^{I J K} A\du{\m\, i}I A
\du{\n\, i} J
A\du{\r \,i} K   \bigg)  + m e \sum_i (\Bar\l{}\du i I \l\du i I) ~~. 
\eqno(3.5) $$   
As usual in any CS theory, the coefficient $~m$~ should be quantized
for a gauge group with non-trivial $~\pi_3\-$homotopy, {\it e.g.},
$$m = \fracmm n{8\pi} ~~~~ (n = \pm 1,~\pm 2, ~\cdots)~~.  
\eqno(3.6) $$  

A key equation useful for the invariance check of (3.5) is
the arbitrary variation of the field strength:
$$ \d F\du{\m\n\, i} I = D_\m (\d A\du{\n \, i}I ) - 
D_\n (\d A\du{\m \, i}I ) ~~, 
\eqno(3.7) $$ 
where the covariant derivative $~D_\m$~ for an arbitrary vector
$~V\du{\m\,i}I$~ with the index $~{\scst i}$~ is defined by 
$$ D_\m V\du{\n\, i} I \equiv \partial_\m V\du{\n\,i} I + f^{I J K} A\du
{\m \, i} J V\du{\n\, i} K - {\scst\left\{ {\r\atop{\m\n}} \right\}} V\du
{\r\, i}I ~~.  
\eqno(3.8) $$      
Here the absence of the symbol $~\sum_i$~ for the second term implies the 
index $~{\scst i}$~ is {\it not} summed.   Relevantly the gauge covariance 
of the field strength under our gauge transformation 
$$\d_{\rm G} A\du{\m\, i} I = D_\m \L\du i I ~~, 
\eqno(3.9) $$ 
has the desirable form:
$$ \d_{\rm G} F\du{\m\n\, i} I = - f^{I J K} \L\du i J F\du{\m\n\, i} K ~~. 
\eqno(3.10) $$     
Again there is no summation over $~{\scst i}$~ in the r.h.s.  

The closure of the gauge algebra (3.1) at the local level is also 
essentially the same as the global case, because the field equation $~\l\du
i I=0$~ delete all the {\it on shell} effect with the gravitino field. 
The on-shell closure yields $~\[ \d_Q(\e_1) , \d_Q(\e_2) \] =
\d_P(i\Bar\e_1^A\g^m\e_2^A) + \d_{\rm E}$, where $~\d_{\rm E}$~ now
implies the extra symmetry on the vector fields 
$$\d_{\rm E} A\du{\m \, i}I = e^{-1} \e\du\m{\r\s} \sum_j a_{i j}
F\du{\r\s \, j}I~~, ~~~~( a_{i j} = - a_{j i}) ~~,   
\eqno(3.11) $$             
with the antisymmetric parameters $~a_{i j}$, leaving the CS
lagrangian (3.3) invariant desirably.\footnotew{We can of course simply 
discard these extra symmetry terms, regarding them as the on-shell 
vanishing terms.}  

We can further generalize our system to a product of different gauge groups:
$~G_1\otimes G_2 \otimes \cdots \otimes G_d$, where $~G_i~{\scst
(i~=~1,~\cdots,~d)}$~ are different gauge groups where $~d$~ is exactly the 
same as the dimensions of the $~L$~ and $~R$~ matrices.    
Accordingly (3.2) and (3.5) can be generalized to  

\newpage

$$F\du{\m\n} {I_i} \equiv \partial_\m A\du\n {I_i} -\partial_\n A\du\m {I_i} + 
f^{I_i J_i K_i} A\du\m {J_i} A\du\n {K_i} ~~,  
\eqno(3.12) $$
$$ \Lag_{\aleph_0\rm CS} = \sum_i \left[ \half m_i \e^{\m\n\r} \left(
F\du{\m\n}{I_i} A\du\r{I_i} - \fracm13 f^{I_i J_i K_i} A\du\m{I_i} A\du\n{J_i}
A\du\r{K_i} \right)  + m_i e \left( \Bar\l{}^{I_i} \l^{I_i} \right) \right] ~~, 
\eqno(3.13) $$    
where for a {\it fixed} index $~{\scst i}$, the $~{\scst I_i,~J_i,~\cdots}$~ 
indices serve as dummy indices, and the quantization of the coefficients 
$~m_i$~ can depend on each gauge group $~G_i$.  The $~f^{I_i J_i K_i}$~ is 
the structure constant for $~G_i$.  Note the peculiar role played by the
$~{\scst i}\-$index, which does {\it not} merely represent a product of
groups of $~G_i$, due to the multiple \alephnull supersymmetries (3.1).         
In other words, the superficially simple-looking lagrangian (3.5) 
actually embraces infinitely many supersymmetries as {\it hidden} 
symmetries!

We stress here again the non-trivial feature of the non-Abelian
SCS theory, namely even though the field strength term in
the lagrangian  (3.5) vanishes, the action still has topological meaning due 
to the non-Abelian term, in particular when the gauge group has non-trivial
$~\pi_3\-$homotopy groups.  On top of that, we have established a system with
\alephnull local supersymmetries.

\bigskip\bigskip\bigskip

\centerline{\bf 4.~~Dimensional Reduction to \alephnull Theories}

It is worthwhile to mention the important relationship of the \alephnull SCS
theory with the 4D self-dual supersymmetric Yang-Mills (SDSYM) theory,
which is the consistent background for the $~N=2$~ open superstring
\ref\siegel{W.~Siegel, \pr{47}{93}{2504}.}.  The importance of the SDSYM 
theory is due to the general conjecture that {\it all} the supersymmetric
integrable systems in $~D\le 3$~ are generated by the SDSYM theory
\ref\gkn{S.V.~Ketov, H.~Nishino and S.J.~Gates, Jr., \np{393}{93}{149};
H.~Nishino, S.J.~Gates, Jr., and S.V.~Ketov, \pl{307}{93}{331}; 
S.V.~Ketov, S.J.~Gates, Jr.~and H.~Nishino, \pl{308}{93}{323}.},
which is the ``supersymmetrization'' of the non-supersymmetric conjecture by
M.F\.Atiyah \ref\atiyah{M.F\.Atiyah, unpublished; {\it ``Classical Geometry of
Yang-Mills Fields''}, (Scuola Normale Superiore, Pisa, 1979); 
M.F\.Atiyah and N.J\.Hitchin, {\it ``The Geometry and Dynamics of Magnetic
Monopoles''}, (Princeton Univ.~Press, 1988), R.S\.Ward and R.O\.Wells, Jr., 
{\it Twistor Geometry and Field Theory''}, (Cambridge Univ.~Press, 1970);  
R.S\.Ward, Phil\.Trans\.Roy\.Lond\.{\bf A315} (1985) 451; 
N.J\.Hitchin, Proc\.Lond\.Math\.Soc\.{\bf 55} (1987) 59.}.  As a matter of 
fact, a recent study \gr\ shows that a set of conjectural \alephnull
supersymmetric integrable equations can be embedded into the \alephnull
supersymmetric YM theory in 3D.  Even though ref.~\gr\ suggested that
the 4D SDSYM theory does not seem to generate arbitrary number of
\alephnull  supersymmetries in 3D, we are going to show that there is a
dimensional reduction scheme, such that the 4D SDSYM theory with {\it finite} 
$~N~$ indeed generates {\it infinitely} many supersymmetries.  

Our scheme of dimensional reduction is much similar to the method
used in ref.~\ref\nishinosdcs{H.~Nishino, \mpl{9}{94}{3255}.}, namely we can
think of a torus compactification of ~$N=4$~ SDSYM \gkn\
in 4D on $~\IR^3\otimes S^1$.  Here instead of directly using the $~N=4$~ 
SDSYM in 4D, we use a $~N=1$~ SDSYM in 4D \gkn\ obtained from the former by some
truncation of fields, and its action is     
$$\li{I\,_{\rm SDSYM}^{N=1} & = \int d^4 x \int d^2 \Hat\theta ~ \Hat 
\L^{\hatm\, I} \Hat W\du{\hatm} I  
&(4.1)  \cr  
& = \int d^4 x \,\bigg[- \half \Hat G^{\hatm\hatn
\, I}\left( \Hat F\du{\hatm\hatn}I - \half\Hat\e\du{\hatm\hatn}{\hatr\hats} 
\Hat F\du{\hatr\hats}I \right)   + i \Hat\r^{\,\hat\a\, I}
(\Hat\G^{\hatm})\du{\hat\a}{\hat{\Dot\b}} \Hat D_{\hatm}
\Hat{\Bar\l}\du{\hat{\Dot\b}}I + \Hat\varphi^I \Hat D^I + \Hat\psi
^{\hatm\, I} \Hat\l\du{\hatm} I \bigg] ~~. \cr} $$
As usual \ng, all the {\it hatted} quantities and indices refer to 4D.  

We now apply exactly the same dimensional reduction scheme as eqs.~(3.3)
through (3.13) in ref.~\ng, {\it except} that now we introduce multiple gauge
groups $~G_{\rm total} = G \otimes G \otimes \cdots \otimes G = G^d$~ with 
the superfields  $~\Hat A_{\hat A}(\Hat z)^{I_i}~{\scst (i~=~1,~2,~\cdots,
~d)}$, where $~{\scst I_i}$~ is for the adjoint representation for the
$~i\-$th gauge group in $~G^d$.  However, we can equivalently use 
these  $~{\scst i}\-$indices
as $~\Hat A_{\hat A\,i}(\Hat z)^I~$, distinguishing the superfields.  By this
prescription for the torus compactification on $~\IR^3\otimes S^1$~, 
we get the action  
$$ \eqalign{ I\, {}_{\rm SDSYM}^{N=1}
~~{\buildrel{^{\rm DR}} \over\longrightarrow}~~   
& \, I\,{}_{\rm SCS}^{N=1} \cr
& \, = \half m \int d^3 x  \int d^2 \theta
\left[ A^\a{}_i(z) W_\a{}_i (z) - \fracmm i 6 (\g^m)^{\b\g} A\du{\b i} I (z) 
A\du{\g i} J (z) A\du{m i} K (z) \, \right] \cr   
& = \int d^3 x \, \sum_i \bigg[ \, \half m \e^{\m\n\r} \left( F\du{\m\n\,i}I
A\du{\r\,i}I - \fracm1 3 f^{I J K} A\du{\m\,i}I A\du{\n\,i} J A\du{\r\,i}K
\right)  + m {\bar\l}_i{}^I \l_i{}^I \, \bigg]~~, \cr } 
\eqno(4.2) $$  
which is nothing but our \alephnull SCS (3.5)!  

When there were no $~{\scst i} \-$summation, this action would be just
an ~$N=1$~ SCS theory in 3D.  However, due to this $~{\scst
i}\-$summation, the system has more {\it hidden} supersymmetries than expected,
promoted to \alephnull supersymmetries under (3.1).
The important point here is that even though we have originally $~N=1$~
supersymmetry from the dimensional reduction, we have ended up with 
hidden promoted supersymmetries of arbitrary number.  It is due to
the {\it on-shell} supersymmetry in the system that such promotions
are possible.  

Before concluding, we mention a similar dimensional reduction/truncation for 
the \alephnull supergravity.  For the \alephnull supergravity we use
the usual $~N=1$~ supergravity in 4D \pvn\ instead of SDSYM 
as the original theory, 
and we perform the dimensional reduction/truncation on $~\IR^3\otimes S^1$. 
The $~N=1$~ supergravity in 4D has the lagrangian \pvn  
$$\Hat\Lag_{\rm SG}^{4D,\,N=1} = - \fracm1 4\Hat R(\Hat\omega) +
\fracm i2 \Bar{\Hat\psi}_{\hat\m} \, \Hat\g^{\hat\m\hat\n\hat\r} \Hat
D_{\hat\n}(\Hat\omega)\Hat\psi_{\hat\r} ~~.    
\eqno(4.3) $$ 
We use the
first-order formalism \pvn, regarding $~\Hat\omega\du{\hat\m}{\hat r\hat s}$~ as
an  independent variable, in order to simplify our dimensional 
reduction/truncation which is similar to (4.2):  
$$\eqalign{&\Hat\psi_\m (x,y) = {\sqrt
2}\sum_{A=1}^\infty \psi\du\m A (x)  \cos(2\pi A y) ~~, ~~~~ \Hat\psi_3 (x,y)
= 0 ~~, \cr  & \Hat e\du{\m}m = e\du\m m (x) ~~, ~~~~
\Hat e\du{3}{(3)} = 1 ~~, ~~~~\Hat e \du 3 m = 0 ~~, ~~~~
\Hat e \du m {(3)} = 0 ~~, \cr 
&\Hat\omega\du{\m}{m n} = \omega\du\m{m n}(x) ~~.  
~~~~ \Hat\omega\du{3}{m n} = 0~~,~~~~ \Hat\omega\du{\m}{(3) n} = 0 ~~. \cr } 
\eqno(4.4) $$ 
Here $~\Hat x^\m = x^\m~{\scst (\m~=~0,~1,~2)}$~ and $~\Hat x^3 = y~ (0\le y
<1)$~ coordinates represent the 3D part and the ``extra'' coordinate in
4D for the reduction/truncation.  Notice that we setup the
$~y\-$dependence only for the gravitino field.  

Performing now our dimensional reduction/truncation, we get the \alephnull 
supergravity in the first-order formalism in 3D: 
$$\eqalign{I_{\rm SG}^{4D,\,N=1} = &\, \int d^3 x \int_0^1 d y ~ 
\Lag_{\rm SG}^{4D,\,N=1}  \cr 
= &\, \int d^3 x ~ \Lag_{\aleph_0\rm SG}^{3D} \cr 
= &\, \int d^3 x ~ \left[\, - \fracm1 4 R(\omega) + \sum_A\fracm i 2 
\Bar\psi\du\m A \g^{\m\n\r} D_\n(\omega) \psi\du\r A \, \right]~~.  \cr}  
\eqno(4.5) $$
Note that all the terms under the $~y\-$integration are always bilinear, since
we are using the first-order formalism, and we have used the relations 
for $~A,~B \in \{ 1,2,\cdots \}$~ like 
$$ 2 \int_0^1 d y ~\cos(2\pi A y) \cos(2\pi B y) = \d_{A B} ~~,~~~~
\int_0^1 d y~ \cos(2\pi A y) \sin(2\pi B y) = 0 ~~.  
\eqno(4.6) $$  
We can also truncate any of $~\psi\du\m A$, so that the summation in (4.5) 
is a finite one from $~{\scst A~=~1}$~ to a finite but arbitrary $~N$.  

An interesting point here is that even though the original gravitino had
{\it finite} degrees of freedom in 4D, it yields an {\it infinite} 
number of gravitini
with {\it infinitely} many supersymmetries in 3D!  In other words, the
\alephnull supersymmetries emerge as hidden symmetries out of our dimensional
reduction/truncation from 4D.  This is possible thanks to the peculiar
property of 3D where a supergravity multiplet has no
physical degree of freedoms.\footnotew{If we try a similar
procedure in a higher-dimensional supergravities, such as from
11D to 10D, we lose consistent supersymmetries in 10D after the
dimensional reduction.  This is because on-shell \alephnull supergravity
is {\it not} possible in 10D.  To put it differently, the
dimensional reduction scheme (4.4) does not maintain
supersymmetries in general higher-dimensions.}

\bigskip\bigskip\bigskip

\centerline{\bf 5.~~\alephnull BF Theories Coupled to SG}

It is now straightforward to consider another important theory in 3D,
namely BF theory.  For this purpose we need two independent
Abelian vector multiplets $~(A_{\m\, i} , \l_i)$~ and $~(B_{\m\,
i}, \chi_i)$, with the supertranslation rules
$$\eqalign{& \d_Q A_{\m \, i } = + \fracm1{2{\sqrt2}}
\sum_j (L^A)_{i j} \left(\Bar\e{}^A \g_\m \l_j \right) ~~, \cr 
&\d_Q \l_i = - \fracm1{2{\sqrt2}} \sum_j (R^A)_{i j} ( \g^{\m\n}
\e^A) \Hat F_{\m\n\, j} + \fracm i 2 \l_i \left( \Bar\e{}^A \g^\m \psi_\m{}
^A \right)  ~~, \cr 
& \d_Q B_{\m \, i} = + \fracm 1{2{\sqrt2}}
\sum_j (L^A)_{i j} \left(\Bar\e{}^A \g_\m \chi\low j \right) ~~, \cr 
&\d_Q \chi\low i = - \fracm 1{2{\sqrt 2}} \sum_j (R^A)_{i j} ( \g^{\m\n}
\e^A) \Hat G_{\m\n\, j} + \fracm i 2 \chi\low i \left(\Bar\e{}^A \g^\m 
\psi_\m{}^A \right)  ~~, \cr }
\eqno(5.1) $$ 
where we are considering only Abelian vector multiplet.  The field strengths
are defined by 
$$ F_{\m\n\,i} \equiv \partial_\m A_{\n \, i} - \partial_\n A_{\m \, i}~~,
~~~~G_{\m\n\,i} \equiv \partial_\m B_{\n \, i} - \partial_\n B_{\m \, i}~~,    
\eqno(5.2) $$
and their {\it hats} denote the supercovariantizations as (3.4).   
The structure of this multiplet is similar to (3.1) except for the
terms with gravitini, which depend on the structure of the lagrangian, 
like the auxiliary-field terms vanishing by the field equations.   

The \alephnull BF lagrangian is given by 
$$\Lag_{\aleph_0\rm B F} =  \half \e^{\m\n\r} \sum_i B_{\r\, i} 
F_{\m\n\, i} + e \sum_i \bigg( \Bar\l{}_i \chi\low i \bigg) ~~. 
\eqno(5.3) $$ 
The invariance check of (5.3) is easy, because the only effect by local
supersymmetry is the $~\psi \chi\l\-$terms arising from the second term, 
which cancel by themselves by the help of the gravitino-dependent terms in
(5.1).  

We give also an alternative \alephnull BF theory based on the on-shell 
3D, \alephnull-supersymmetric vector multiplet with field content
~$(A_\m , \,  {\cal B}_i {}^j ,  \, \l_{\a \, {\rm I}} , \, {\Hat
\l}_{\a \, \hat k} {}^k )$. The variations of these fields
are given by \gr
$$ \eqalign{ {~~~~~~}
\d_Q A_\m &=~ - \, i \e^{\a \, {\rm I}} \,  (\g_\m)_{\a\b}
\, \l^\b{}_{\rm I}  ~~~, \cr
\d_Q {\cal B}_i {}^j &=~ \e^{\a \, {\rm I}}  \left[ ~ \left( f_{{
\rm I} \, {\rm J}} \right)_i {}^j \,  \l_{\a \, {\rm J}}  \, + \,
\left( {\rm L}_{\rm I}  \right)_i {}^{\hat k} \, {\Hat \l}_{\a \,
\hat k} {}^j ~ \right] ~~~, \cr                                  
\d_Q \l_{\a \, {\rm I}} &=~ i \, \e^{\b J}(\g^\m)_{\a\b}
\left[ ~ \frac 12 \d_{\rm {I \, J}} \e\du\m{\r\s} F_{\r\s} (A)  ~-~
{\rm d}^{-1} \, \left( f_{{\rm I} \, {\rm J}} \right)_{i} {}^{j} ( \,
\pa_\m {\cal B}_j {}^i \,) ~ \right] ~~~, \cr
\d_Q {\Hat \l}_{\a \, \hat k} {}^k &=~  i \, \e^{\b \, {\rm J}} (
\g^\m)_{\a \b}  \left[ ~ ({\rm R}_{\rm J} )_{\hat k} {}^i ( \, \pa_\m
{\cal B}_i {}^k \,)  ~-~ {\rm d}^{-1} \, ({\rm R}_{\rm I} )_{\hat k}
{}^k \left( f_{{\rm I} \, {\rm J}} \right)_{i} {}^{j} ( \, \pa_\m {\cal
B}_j {}^i \,) ~ \right] ~~~. \cr }
\eqno(5.4)$$
The existence of this on-shell supersymmetric representation suggests
that there is another 3D, \alephnull-supersymmetric vector multiplet
that is dual to the one above in such a way that a 3D, \alephnull
-supersymmetric BF action exists.  This purported theory in the
special case of ~$N = 4$~ has already been constructed 
\ref\brooksgates{R.~Brooks and  S.J.~Gates, Jr., \np{432}{94}{205}}.  
In the following, we generalize this result to all values of ~$N$.

The first step in our generalization is to note that the fields of
our expected on-shell dual 3D, \alephnull-supersymmetric vector
multiplet can be written in the form $(B_\m  , \, \b_{\a
\, {\rm I}} , \, {\Hat \b}_{\a \, k} {}^{\hat k}
, \,  {\rm d}_i {}^j)$.  We want this
vector multiplet to be dual to the one above in the
sense that its components can appear in an action that contains
the usual BF coupling between $A_\m$ and $B_\m$.  For this purpose we write,
$$ {\cal L}_{\aleph_0 {\rm B} {\rm F}}'  ~=~ \frac 12 \e^{\m\n\r}
B_\m F_{\n\r} (A) ~-~
\b^{\a \, {\rm I}} \l_{\a \, {\rm I}} ~-~ {\rm d}^{-1}
{\Hat \b}^{\a}{}_{ k} {}^{\hat k} {\Hat \l}_{\a \,
\hat k} {}^k ~+~ {\rm d}^{-1} {\rm d}_i {}^j
{\cal B}_j {}^i  ~~~,  
\eqno(5.5) $$
and in such a way that the action is a supersymmetric invariant. The
requirement that this action is left invariant under a supersymmetry
variation can be used to determine the appropriate variations
for $(B_\m  , \, \b_{\a \, {\rm I}} , \, {\Hat \b}_{\a \, k} {}^{\hat k}
, \,  {\rm d}_i {}^j)$
$$ \eqalign{ {~~~~~~}
\d_Q B_\m &=~ - \, i \e^{\a \, {\rm I}} \,  (\g_\m )_{\a \b}
\, \b^{\b}{}_{\rm I}  ~~~, \cr
\d_Q \b_{\a \, {\rm I}} &=~  \e^{\b \, {\rm J}} \left[ ~ i  \frac 12 \,
(\g^\m)_{\a \b} \d_{\rm {I \, J}} \e_\m {}^{\n\r} F_{\n\r} (B) ~-~
{\rm d}^{-1} \, \left( f_{{\rm I} \, {\rm J}} \right)_{i} {}^{j}
{\rm d}_j {}^i ~ \right] ~~~, \cr
\d_Q {\Hat \b}_{\a \, k} {}^{\hat k} &=~ \e_{\a}{}^{\rm I}
\left[ ~{\rm d}_k {}^j \left( {\rm L}_{\rm I}  \right)_j {}^{\hat k}
~+~ {\rm d}^{-1}   \left( {\rm L}_{\rm J}  \right)_k  {}^{\hat k}
\left( f_{{\rm I} \, {\rm J}} \right)_{i} {}^{j} {\rm d}_j {}^i ~
\right]
~~~, \cr
\d_Q ~ {\rm d}_i {}^j &=~ i \e^{\a \, {\rm I}}  (\g^\m )_{\a \g}
\pa_\m \, \left[ ~
\b^{\g \, {\rm J}} \left( f_{{\rm I} \, {\rm J}} \right)_{i} {}^{j}
~+~ {\Hat \b}^{\g} {}_{ i} {}^{\hat k} ({\rm R}_{\rm I} )_{\hat k}
{}^j ~ \right]  ~~~.}
\eqno(5.6)$$

In closing this section, we note that the existence of the this
3D, \alephnull-supersymmetric BF action together with the
existence of 3D, \alephnull-supersymmetric scalar multiplets
should naturally lead to 3D, \alephnull-supersymmetric anyonic
models. A further challenge will be to investigate the further
existence of 3D, \alephnull-supersymmetric CS actions that 
possess anyonic extensions.  Finally we note that given the action 
in (5.5), we expect a further duality transformation exists that
permits the last term to be replaced by BF-type terms.  (See the
two different $~N = 2$~ theories of \ng.)

\bigskip\bigskip\bigskip

\centerline{\bf 6.~~\alephnull Supergravity Liouville Theory in 2D}

We have so far discussed \alephnull supergravity theories only in 3D.  As an
interesting application of such theories, we perform the dimensional
reductions of them to get 2D \alephnull theories.  A typical example we give
here is \alephnull supergravity  Liouville theory.  In this paper, we skip 
all the details of the dimensional reduction but only the final results 
which will be of more interest for other applications.    

Our metric in 2D is ~$ (\eta_{m n} ) = ( \eta_{0 0}, \eta_{11} ) \equiv (+1,-1)
$, and accordingly we have $~\g^{m n} = + \eta^{m n} + \e^{m n} \g_5$.  
Our multiplets are the \alephnull supergravity 
$~(e\du\m m,\psi\du\m A)$~ and a dilaton multiplet $~(\varphi, \chi^A)$.  Here
the indices $~{\scst A,~B,~\cdots~=~1,~\cdots,~N\rightarrow\infty}$~ are for
the $~N\rightarrow\infty\-$extended supersymmetries. 
The invariant lagrangian $~\Lag_0 + \Lag_g$~ for our
\alephnull-extended supergravity coupled to \alephnull Liouville theory is given
by  
$$ \eqalign{ e^{-1}\Lag_0 = \, & +\varphi R - 2e^{-1} \e^{\m\n}
\left(\Bar\chi\g_5 {\cal R}_{\m\n}  \right) + \half (\partial_\m\varphi)^2 +
\fracm i 2 (\Bar\chi\g^\m D_\m \chi) \cr 
&  - \half (\Bar\psi_\m \g^\n\g^\m \chi) (\partial_\n \varphi + \Hat
D_\n \varphi) ~~, \cr 
e^{-1}\Lag_g = \, & - 8g^2 e^\varphi + 4i g e^{\varphi/2} (\Bar\psi_\m\g^\m
\chi) - 8 g e^{-1} e^{\varphi /2} \e^{\m\n} (\Bar\psi_\m \g_5 \psi_\n) + g
e^{\varphi/2} (\Bar\chi\chi) ~~, \cr }  
\eqno(6.1) $$
where $~\left(\Bar\chi\g_5 {\cal R}_{\m\n} \right) \equiv \left( \Bar\chi{}^A
\g_5{\cal R}_{\m\n}{}^A \right)$, {\it etc.}  Eq.~(6.1) is 
invariant under the supersymmetry 
$$ \eqalign{ & \d_Q e\du\m m = - 2 i ( \Bar\e{}^A \g^m \psi\du \m A ) \equiv -
2i (\Bar\e\g^m\psi_\m) ~~, \cr 
& \d_Q  \psi\du\m A = \partial_\m \e^A + \half \omega_\m\g_5 \e^A - \fracm i
{32} 
\g_\m \g^\n \e^B (\Bar\chi{}^A \g_\n\chi^B) \cr 
& ~~~~~ ~~~ + \fracm1 4 \g^\n \chi^{\[ A} ( \Bar\e{}^{B\]} \g_\n \psi\du\m
B) + \fracm 1 4 \g^\n \chi^{\[A|} (\Bar\e{}^B \g^\n \psi\du\m{|B\]} ) ~~\cr 
& \d_Q  \varphi = ( \Bar\e{}^A \chi^A ) \equiv (\Bar\e \chi) ~~, \cr 
& \d_Q \chi^A = - i \g^\m \e^A \Hat D_\m \varphi + 4 g e^{\varphi/2}  \e^A -
\fracm 1 8 \g^\m \e^B (\Bar\chi{}^A \g_\m \chi^B) ~~,  \cr 
& \d_Q \omega_\m \equiv +i e^{-1} \e^{\r\s} (\Bar\e \g_\m{\cal R}_{\r\s}) 
~~\cr }  
\eqno(6.2) $$ 
Here we have defined 
$$\eqalign{ & R \equiv + 2 e^{-1} \e^{\m\n} \partial_\m\omega_\n  ~~, ~~~~ 
{\cal R}_{\m\n}{}^A \equiv D_\m\psi_\n{}^A - D_\n \psi_\m{}^A ~~, \cr 
& D_\m \e^A \equiv \partial_\m \e^A + \half \omega_\m \g_5 \e^A ~~,
~~~~\omega_\m \equiv - e^{-1} \e^{\r\s} \left( e\du\m m \partial_\r e_{\s m} + i
\Bar\psi_\r \g_\m \psi_\s \right) ~~, \cr  }  
\eqno(6.3) $$ 
The constant $~g$~ controls the potential term with the exponential function of
the dilaton $~\varphi$ as usual in a Liouville theory in 2D.  Note that in the
case of simple supersymmetry ($N=1$), all the fermionic bilinear terms in (6.2)
disappear.      

This invariant lagrangian and the supersymmetry transformation rules
are fixed by the usual method, namely cancelling the derivative on the
parameter in the supersymmetric transformation of fermionic field equations
by adding fermionic quartic terms to the lagrangian and the fermionic
bilinear terms to the transformation rules for fermions.  
Notice that there is no explicit quartic terms in the lagrangian 
or bilinear fermions in the
supertransformations of fermions when $~g=0$.

\bigskip\bigskip

\bigskip\bigskip\bigskip

\centerline{\bf 7.~~Concluding Remarks}

In our paper we have presented an amusing result that  
on-shell Poincar\'e supergravity in 3D can be extended up to
$~N=\infty$~ with the minimal field content only with the dreibein and the
gravitini which we call minimal \alephnull supergravity.  This system can be
further coupled to CS as well as BF theories.  This result is also consistent 
with the recent on-shell results in \howe, when central charges are
present.        

We have overcome the previous difficulty with supersymmetrizing CS theories 
beyond $~N=4$~ \ng, by introducing the $~L$~ and $~R\-$matrices.  
In the on-shell formulation, all the vanishing terms by field equations can 
be also understood as the extra symmetries \ng.  This prescription seems
possible only in 3D due to the special property of the dreibein and
gravitini which are essentially non-physical, making the on-shell closure of
the gauge algebra simple.  The usual upper limit for the number of
supersymmetries does not apply to 3D because of these non-physical
dreibein and gravitini.  If we try to couple supergravity to
``physical'' scalar multiplet with the usual kinetic terms, we encounter an
obstruction against consistent Noether couplings.  This fact also shows the
importance of the non-physical property of all the  fields as the key
feature of the system.  In the context of conformal  supergravity \ng, a
similar phenomenon in 3D has been already encountered, in which
arbitrary number of supersymmetries up to infinity are allowed.  To our
knowledge, however, our system is the first example with \alephnull
Poincar\'e supersymmetries, including also the ``matter'' multiplets.  

We can re-interpret our results for \alephnull SCS theories as follows. 
Reviewing (3.5), we can regard it just as an $~N=1$~ SCS theory for a product
of the same Yang-Mills gauge group: $~G^n \equiv G\otimes G\otimes\cdots\otimes
G$, and we are re-labeling $~A\du\m{I_i}$~ as $~A\du{\m i}I$~ {\it etc.}~as
given in \ng.  However, such a system has {\it hidden} enlarged supersymmetries
promoted to $~N=n$, where ~$n$~ coincides that in $~G^n$, and the enlarged
supersymmetry is characterized by the $~L$~ and $~R$~ matrices in (3.2). 
As a matter of fact, this new observation has overcome the previous difficulty 
to go beyond $~N=4$~ in ref.~\ng.          
  
Our result in three-dimensional systems is natural also from the
viewpoint of strong-weak coupling duality suggested by Witten \wittendual.  
The appearance of arbitrarily many supersymmetric gravitino fields in
three-dimensions may be understood as a reminiscent of ``dimensional
reduction''\footnotew{This is not ``reduction'' in its strict sense, because 
the radius of the $~S^1$~ will be infinity instead of zero.} of some
four-dimensional theory, in which there is a finite number of gravitini.  
Upon a compactification of such a theory on $~\IR^3 \times S^1$~ with an
$~S^1$~ of an infinitely large radius will yield a set of infinitely many
gravitini.  Thus from the 4D viewpoint this limit is understood as the
weak coupling, while from the 3D viewpoint this limit can be shown to 
be equivalent to taking the string coupling $~\l\rightarrow\infty$~ limit
\wittendual.  In an ordinary ``reduction'' into other dimensions, this
usually fails due to the inconsistency with the couplings of gravitini.  The
special feature of the three-dimension is that the gravitini are no longer
physical fields, but rather non-propagating fields, that enable us to
construct couplings to CS and BF theories which also have only non-physical
fields.          

This feature can be more elucidated, when we compare it with the conformal 
supergravity in 2D \ref\bns{{\it See e.g.,} E.~Bergshoeff,  H.~Nishino and 
E.~Sezgin, \pl{165}{86}{141}.}.  In the latter, we do not have any field 
equations such as the vanishing gravitino field strengths, so that even
on-shell closure of gauge algebra was not manifest, and therefore we needed
more field to realize its closure off-shell.  It is very
peculiar to the 3D theories, where lagrangians produce vanishing field
strengths, which make the closure of the gauge algebra manifest.  

The importance of 3D systems of this type is being increasingly recognized. 
Most recently, it has been suggested \ref\wittendual{E.~Witten, {\it ``Strong
Coupling Expansion of Calabi-Yau Compactification''}, preprint, IASSNS-HEP-96-08
(Feb. 1996) hep-th/9602070.} that the appearance of the non-perturbative 
potential of 4D heterotic string theory is governed by 3D physics.
An intriguing question to ask is whether there exists an
``$\large{\aleph_0}$~ string-like'' theory that incorporates all of such
theories.

\bigskip\bigskip\bigskip

The authors are grateful to R.~Brooks who first drew their attention to
the importance of BF theories coupled to extended supergravity.      

\bigskip\bigskip\bigskip

{\bf {Added Note in Proof}}

After the completion of this paper, the authors became aware of a work
(Devchand and Ogievetsky, ``Interacting Fields of Arbitrary Spin
and $~N>4$~ Supersymmetric Self-Dual Yang-Mills Equations'' ICTP 
preprint IC/96/88, hep-th/9606027) which reports to prove the
existence of \alephnull supersymmetric self-dual Yang-Mills theory in 
4D. This provides independent and additional support for the
new class of integrable systems proposed in reference [1].

\vfill\eject 

\footatend\vfill\supereject\immediate\closeout\rfile\writestoppt
\baselineskip=14pt\centerline{{\bf References}}\bigskip{\frenchspacing%
\parindent=20pt\escapechar=` \input refs.tmp\vfill\eject}\nonfrenchspacing

\end{document}